\documentclass[12pt,thmsa]{article}
\usepackage{amsfonts}
\usepackage{graphicx}
\usepackage{epsfig}
\usepackage{graphics}

\makeatletter
\newcommand{\row}[1]%
{\mathord{\buildrel{\lower3pt%
\hbox{$\scriptscriptstyle\rightarrow$}}\over #1}}

\newcommand{\dyadic}[1]{\mathord{\dyadic@rrow{#1}}}
\newcommand{\dyadic@rrow}[1]{
\begin{picture}(12,12)(-1,0)
\put(-2,12){\makebox(0,0)[t]{$\scriptscriptstyle\downarrow$}}
\put(-2,12){\makebox(0,0)[l]{$\scriptscriptstyle\longrightarrow$}}
\put(5,0){\makebox(0,0)[b]{$#1$}}
\end{picture}
}

\newcommand{\ket}[1]{\bigl| #1 \bigr\rangle}
\newcommand{\expect}[1]{\left\langle #1 \right\rangle}
\begin{document}

\begin{center}
{\Large  Abrupt and gradual changes of information through the
Kane solid state computer}

~

\small{Nasser Metwally}

{\footnotesize Mathematics Department, College of Science, Bahrain
University, 32038 Bahrain }
\end{center}

\begin{abstract}
The susceptibility of the transformed  information  to the filed
and system parameters is investigated for the  Kane solid state
computer. It has been shown, that the field polarization  and the
initial state of the system play the central roles on the abrupt
and gradual quench of the purity and the fidelity. If the field
and the initial state are in different polarizations, then the
purity and the fidelity decrease abruptly, while for the common
polarization the decay is gradual and smooth. For some class of
initial states one can  send the information without any loss.
Therefore, by controlling on the devices one can  increase the
time of safe communication, reduce the amount of exchange
information between the state and its environment and minimize the
purity decrease rate.

\end{abstract}
{\bf Keywords}: Information, Quantum Computer, Fidelity, Entropy

{\bf PACS} 03.65.-w, 03.67.-a, 03.65.Yz
\section{Introduction}
In recent years the ambitions on quantum computer and  its new
technology, has  gained much attention \cite{shor,Mich}. It is a
device for computation that makes  use of quantum mechanical
characterization , such as superposition and entanglement
\cite{Br}. It manipulate two level quantum system called qbit
(quantum bit) that can be represented as an  arbitrary
superposition of $\ket{0}$ and $\ket{1}$ \cite{Deut}. There are
many attempts have been done to produce a scalable quantum
computer. Among these temptats, Nuclear Magnetic Resonance (NMR)
\cite{vand}, but unfortunately the constructed quantum computer
does not work beyond a few tens of qubits \cite{Jones}. The
systems of trapped ions  have been represented as  models of
quantum computers. In these systems one encodes and manipulates
information via long-lived internal states of ions
\cite{Cirac,sor}. Also, As a candidate of scalable solid-state
quantum computer, superconducting circuits with Josephson junction
\cite{Mak} have attracted much attention in recent years. The
perceived advantages of solid state quantum computation are its
scalability and  the low decoherence rate for spin states
\cite{sarma}.

Quantum computation are based on a set of quantum control-NOT
(CNOT) gate performed  between two qubits \cite{Mich}. Due to the
noise these CNOT operations are not performed correctly and
consequently  the quantum computations  are not correctly
performed \cite{kane,Bacon}. Wellard and Hollenberg \cite{Well}
have investigated the CNOT operation of the Kane quantum computer,
where the evaluation of a single qubit is controlled by a voltage.
Sometimes this voltage contains a source of stochastic noise which
leads to  noisy operations. Also the influence of random errors in
external control parameters on the stability of holonomic quantum
computation has been investigated in \cite{Bui}. So, quantum state
which carries the information is too fragile to exist for a long
time unless the system under consideration is perfectly isolated
from its environment. This detrimental effect of decoherence
causes a loss of transmitted information and the loss ofpurity of
its carrier \cite{Murt}.

 This motivates us to unveiling some properties of the
 information's carrier. Among these properties is the purity,
which is  employed for investigating the process of coherence loss
\cite{zur,Jun}, entropy which measures the amount of exchange of
information between the system and its environment \cite{Yan,Hol}
and the  coherent vectors, which give one of the possible
descriptions of N-level quantum state and  allow us to grasp
characteristics of states from a completely geometrical standpoint
\cite{Gen}. Effective transport of quantum information is an
essential element of quantum computation. Since the desired
information is encoded in the carrier state which is subject to an
external noise, so  it is important to evaluate the fidelity of
the transmitted information.

 The paper is organized as
follows: In Sec.$2$, we introduce the model. The dynamics of
purity and entropy  is given in Sec.$3$. Investigation of gradual
and abrupt changes on the coherent vector is the subject of
Sec.$4$. In Sec.$5$, the dynamics of the transmitted information
is investigated. Finally we conclude and discuss our results in
Sec.$6$.

\section{The description of the model}
The system under consideration  consists of spin $\frac{1}{2}$
$^{31}P$ nuclei in a silicon  substrate qubit. This system is used
as the main block of  the solid state quantum computer(SSQC),
which has been proposed by Kane \cite{kane}, based on an array of
individual phosphorus donor atoms embedded in a pure silicon
lattice. Both the nuclear spins of the donors and the spins of the
donor electrons participate in the computation.
 These qubits are  subjected  to a background magnetic field in the $z-$ direction.
 At low energy the effective Hamiltonian  for the nucleus-electron system is  given
by \cite{Well}
\begin{equation}
H_{eff}=B_z(\mu_b\sigma_z-g_{\mu}\mu_n\tau_z)+A\row\sigma\cdot\row\tau,
\end{equation}
where $\row\sigma=(\sigma_x,\sigma_y,\sigma_z)$ and
$\row\tau=(\tau_x,\tau_y,\tau_z)$, are Pauli's operators for the
nucleus and the electron qubits respectively, $\mu_b$ is  the Bohr
magneton, $\mu_n$ the nuclear magneton and $g_b$ the g-factor for
the phosphorous nucleus. The constant $A=\frac{\pi}{3}\mu_bg
g_n\mu_n |\psi_1(0)|^2$, and $\psi_1(0)$ is the state vector of
the electron wave evaluated at the position of the nucleus. If
$A<<2B_z\mu_b$, then  the effective Hamiltonin for the first order
in $A/\mu_bB_z$ as
\begin{equation}\label{eff}
H_{eff}=B_z\gamma\sigma_z,
\end{equation}
where, $B_z$ is a background magnetic field,
$\gamma=\gamma_0+\frac{\eta\hbar v}{B_z}$,
$\gamma_0=-g_n\mu_b-\frac{A}{B_z}$. The parameter $\eta$ describes
the rate of change of the nuclear magnetic resonance frequency of
the qubit as a function on the A gate voltage (for more details
see\cite{Well}). Also, if  the field is in its resonance frequency
i.e. $\omega=\frac{2B_z\gamma}{\hbar}$, then the Hamiltonian of
the field is
\begin{equation}
H(t)=B_{ac}\acute{\gamma}[cos(\omega t+\phi)\sigma_y-\sin(\omega
t+\phi)\sigma_x],
\end{equation}
where $\acute{\gamma}=-g_n\mu_n$ and $B_{ac}$ is the transfer
magnetic field. For $\phi=0$, one gets the Hamiltonian which
describes the qubit in the quantum register that is  a qubit not
undergoing any noise. In the presence of noise the Hamiltonian
(\ref{eff}) is given by:
\begin{equation}
H_{eff}=B_z\zeta(t)\sigma_z,
\end{equation}
where $\zeta(t)$, describes the effective Hamiltonian of the
stochastic fluctuations. If, we average over the noise we find
\begin{equation}\label{Ham}
\frac{d\rho_s}{dt}=-\frac{B_{ac}\acute{\gamma}}{i\hbar}\Bigr[
(\cos\theta  \sigma_y-\sin\theta \sigma_x),\rho_s(0)\Bigl]
-\frac{\epsilon B^2_z}{2\bar
h}\Bigr[\sigma_z,[\sigma_z,\rho_s(0)]\Bigl],
\end{equation}
where, $\epsilon=\sqrt{\frac{\eta\bar h V}{B_z}}$, $\theta$ is the
polarization  angle of the field and $\rho_a(0)$ is the initial
state of the qubit system which carries the information
\cite{nasser},
\begin{equation}\label{in}
\rho_s(0)=\frac{1}{2}(1+s_x(0)\sigma_x+s_y(0)\sigma_y+s_z(0)\sigma_z),
\end{equation}
where $s_i(0)=tr\bigl\{\rho_a(0)\sigma_i\bigl\}, i=x,y$ and $z$
are the Pauli vectors (coherent vectors) \cite{Guntr}. Using
Eq.(\ref{Ham}) and Eq.(\ref{in}), one gets the final density
operator of the evolved system $\rho_s(t)$. As soon as one gets
the density operator, we can investigate all the physical
properties of the state and the fidelity of the transformed
information as we shall see below. In our calculation we set the
scaled time $\tau=B_z^2 t$ and $\kappa$ is the ratio between the
transfer magnetic component $B_{ac}$ and the background magnetic
component of the field $B_z$ where we find that
$\kappa=\frac{4B_{ac}}{B_z^2}.$

\begin{figure}[t!]
  \begin{center}
  \includegraphics[width=17pc,height=12pc]{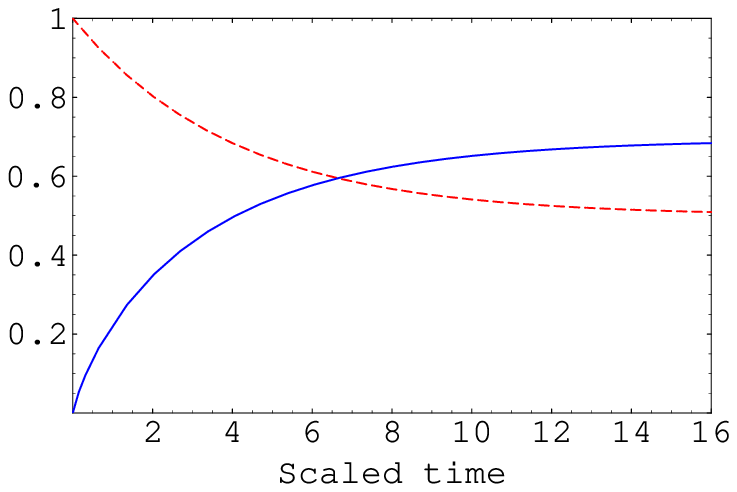}
\includegraphics[width=17pc,height=12pc]{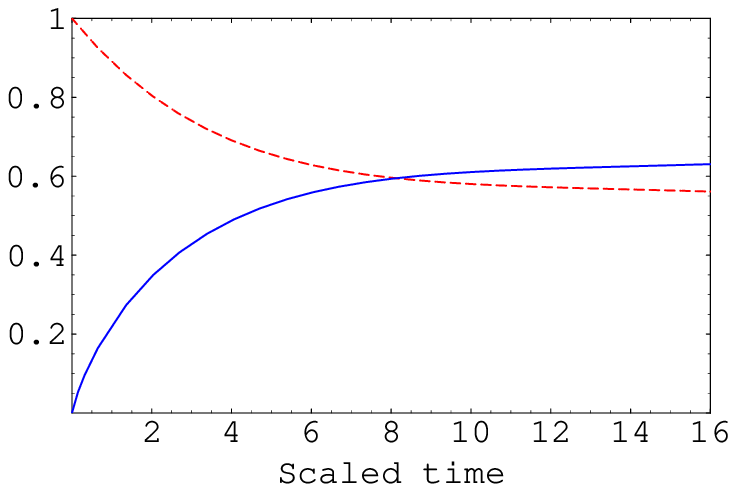}\
\put(-240,125){(a)} \put(-30,125){(b)}\
 \includegraphics[width=17pc,height=12pc]{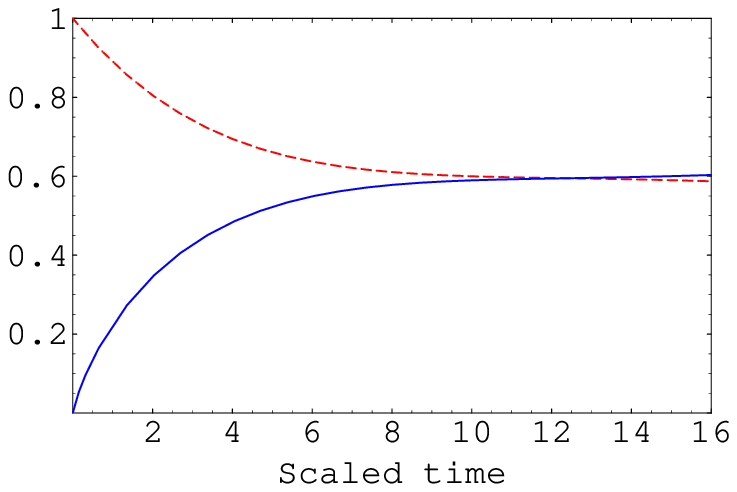}
\includegraphics[width=17pc,height=12pc]{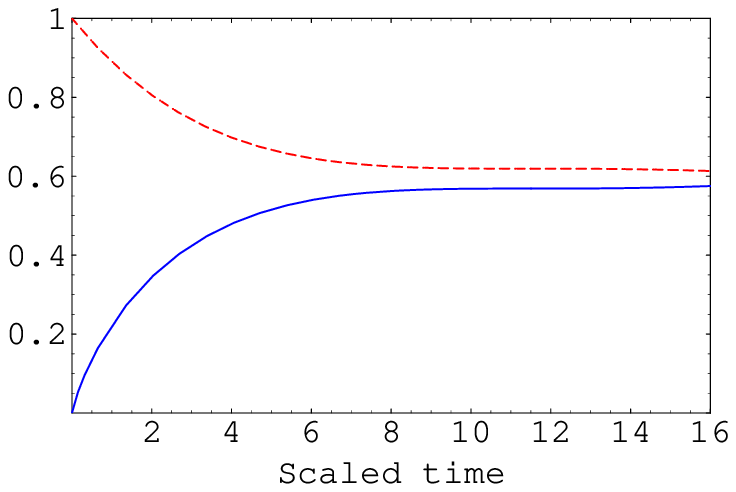}\
\put(-240,125){(c)} \put(-30,125){(d)}\
    \caption{ The dynamics of purity(dot curves) and entropy(solid-curves) for the system is initially prepared with
     $\rho(0)=\frac{1}{2}(1+\sigma_y)$ with $\kappa=\frac{4B_{ac}}{2B_z^2}=0.05$ and  $\theta=0,\frac{\pi}{4},\frac{\pi}{3},
     \frac{\pi}{2}$ for the Figs.$(a),(b),(c$) and $(d)$
     respectively.}
  \end{center}
\end{figure}
\section{Purity and Entropy}
In this section, we investigate some properties of the evolved
state which carries the transfer information. In this context, the
most important  properties are the purity $\mathcal{P}$ and
entropy $\mathcal{E}_n$. To explain the observed behavior of the
purity and the entropy we have to take into account different
classes of initial state setting and different polarization of the
field.

Fig.($1)$, shows the dynamics process of  the purity and the
entropy for different polarizations of the field. {\it Firstly},
let us start with a pure density operator  as an initial state
$\rho(0)=\frac{1}{2}(1+\sigma_y)$ and different polarizations of
the external field. In Fig.(1a), we assume that the field is
polarized in $x$-direction, i.e,  the applied noise is of bit flip
type. It is  evident that, at the scaled time  $\tau=B_z^2 t=0$,
the purity $\mathcal{P}$ is maximum and equals $1$. Then as time
goes on the purity decreases abruptly and reaches  its minimum
value $\mathcal{P}\simeq 0.5$ then it becomes a constant. At this
instance of time  the pure state turns into a completely mixed.
 Also, this figure shows that initially the entropy of the system is zero. As
time increases, the entropy rapidly increases and for late times,
it goes to a constant ($\mathcal{E}_n\simeq0.7)$. In this case
there is no more exchange information  between the state which
carries the information and the environment.

In Fig.(1b), the polarization  angle of the field is
$\frac{\pi}{4}$. In this case, the  rate of purity loss  is
smaller than that depicted in Fig.(1a). Also it quenches gradually
and takes more time to become constant.  Also, the rate  of
entropy increasing is small compared with the corresponding one in
Fig.(1a). Consequently the rate of exchange of information between
the state and the environment decreases. As one increases the
polarized angle of the field, $\theta$  the purity decreases
smoothly, gradually  and  the rate of purity loss is smaller
compared with small  $\theta$. Conversely concerning the entropy
we can notice that the rate of increasing $\mathcal{E}_n$, is
small and hence the exchange of information decreases. This
behavior  can be seen in Fig.(1c) and Fig.(1d), where the
polarization angle of the field $\theta=\frac{\pi}{3},
\frac{\pi}{2}$ respectively.

\begin{figure}[b!]
  \begin{center}
\includegraphics[width=17pc,height=12pc]{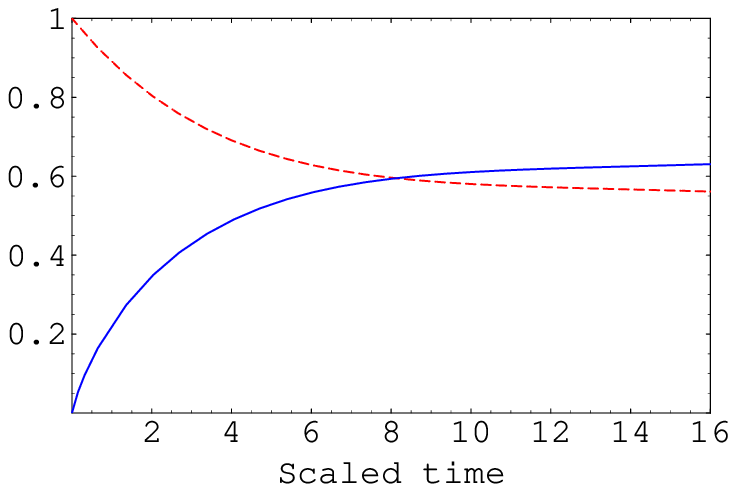}
\includegraphics[width=17pc,height=12pc]{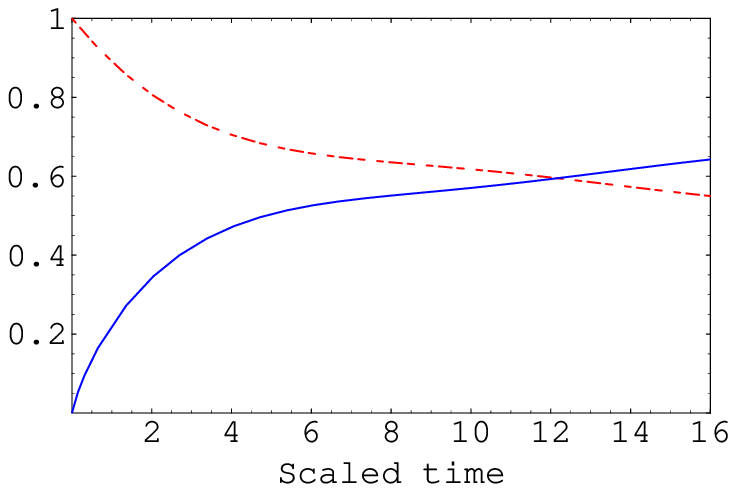}\
\put(-240,125){(a)} \put(-30,125){(b)}\
\includegraphics[width=17pc,height=12pc]{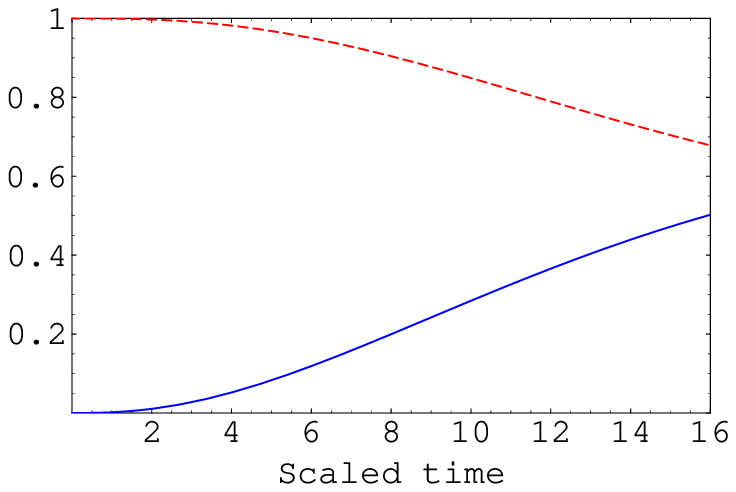}
\includegraphics[width=17pc,height=12pc]{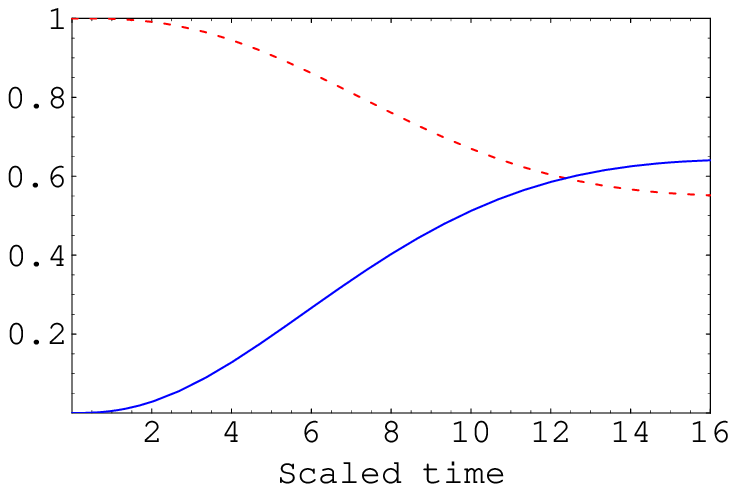}
\put(-240,125){(c)} \put(-30,125){(d)}\
    \caption{ The same as in Fig.(1) but the initial state of the
    system is $\rho_a(0)=\frac{1}{2}(1+\sigma_x)$ for
    Figs.$(a,b)$, with  $\kappa=0.05,0.09$ respectively and
     $\rho_a(0)=\frac{1}{2}(1+\sigma_z)$, for Figs.$(c,d)$
     and $\kappa =0.05,0.09$ respectively.}
  \end{center}
\end{figure}

In Fig.(2), we investigate the effect of the strength of the field
parameter on the dynamics of the purity and the entropy. In
Fig.$(2a)$ and $(2b)$, we start with an initial state completely
polarized in $x$-direction i.e $\rho=\frac{1}{2}(1+\sigma_x)$
while the polarization angle of the field $\theta=\frac{\pi}{4}$.
The behavior of purity and  entropy is similar  as that shown in
Fig.$(1)$. This is clear from Fig.(2a), where we set the value of
the field strength $\kappa=\frac{4B_{ac}}{2B_z^2}=0.05$.  In
Fig.(2b) we increase the value of  $\kappa$, the rate of purity
loss is increased ($\mathcal{P}$ decreases more ), while entropy
increases more.

Finally, we enclose this section with a  system initially prepared
as $\rho(0)=\frac{1}{2}(1+\sigma_z)$ while the polarization angle
of the field, $\theta=\frac{\pi}{4}$. The dynamics of the purity
and entropy for this class of initial states is shown in
Fig.$(2c)$ and Fig.$(2d)$. One can notice that at the first moment
of interaction the purity is almost maximum $(\mathcal{P}=1)$ and
the entropy $\mathcal{E}=0$. This means that for $\tau\leq 2.5$,
one can send the information safely and without any loss. On the
other hand through this time there is no  exchange of information
between the state and  the environment. As time goes on purity
decrease and the entropy increases  smoothly and gradually. As one
increases  the strength  of the field parameter i. e,
$\kappa=0.09$ the change in the purity and entropy  happen
abruptly and much faster.

From our previous discussion, one may say that the dynamics of the
purity and the  entropy of the evolved state  depends on the type
of noise (phase-bit or bit) flip and the polarization of the
initial state. If the noise and the initial state are polarized in
the same direction, then the change in both $\mathcal{P}$ and
$\mathcal{E}_n$ is  gradual. On the other side, this change
becomes  abrupt if  both of the initial state and the field are
polarized in different directions. One of the most appreciated
choices for  generating  the carrier of information is $z-$
direction. In this case the survival purity is shown for a long
time and the entropy increases very slow.

\section{Coherent vector's dynamics}
In this context, one of the most important phenomena of the
evolved state is the dynamics of the amplitude of the coherent
vector, $|s|=\sqrt{s_1^2+s_2^2+s_3^2}$. It is known that a state
can be characterized by expectation values of   $\expect{s_i}$
which are directly observed in experiments. Any density matrix in
2-level systems turns out to be characterized uniquely by a
3-dimensional real vector where the length satisfies $|s|\leq 1$.
Therefore, one  define the Bloch-vector space
$B(R_3):\{s=(s_1,s_2,s_3)\in\Re^3, |s|\leq 1\}$  as a ball with
radius 1. The  surface of the ball  corresponds to the set of pure
states and its inside to mixed states.

Our aim in this subsection to quantify the rate  shrinking of the
amplitude value of the coherent vector. So in Fig.(3) and Fig.(4),
we discuss this phenomenon for different classes of states and
field parameters.
\begin{figure}[b!]
  \begin{center}
  \includegraphics[width=17pc,height=15pc]{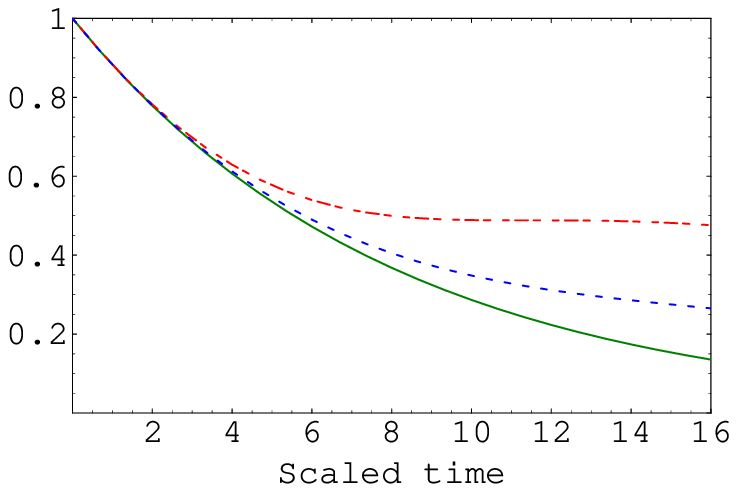}
\includegraphics[width=17pc,height=15pc]{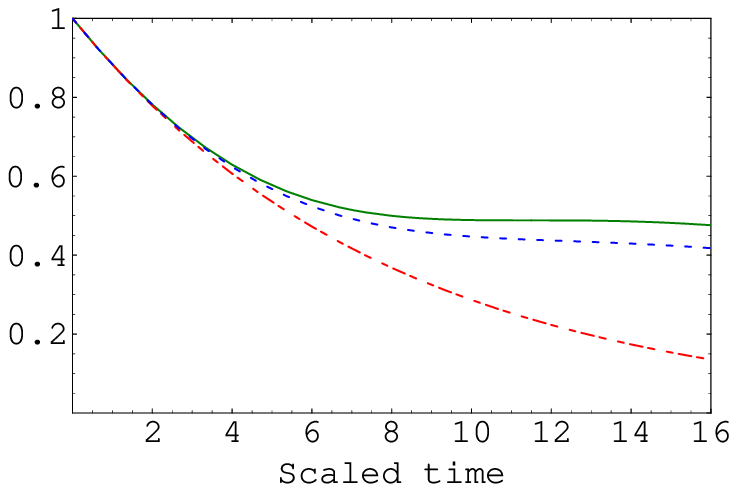}
\put(-240,160){(a)} \put(-30,160){(b)}
    \caption{The dynamics of the amplitude value of the coherent vector $|s|$
    (a) $\rho_{(0)}=\frac{1}{2}(1+\sigma_x)$
     (b) $ \rho_{(0)}=\frac{1}{2}(1+\sigma_y)$
        The solid, dashed and dot  curves for $\theta=\frac{\pi}{2},0,
        \frac{\pi}{3}$ and  $\kappa =0.05$
      respectively.}
  \end{center}
\end{figure}
In Fig.(3a), the initial state is prepared in the pure state
$\rho_s(0)=\frac{1}{2}(1+\sigma_x)$, the field strength
$\kappa=0.05$ and different polarization of the field are
considered. From this figure it is evident that, as soon as the
interaction starts $\tau>0$, the coherent vector decays abruptly.
Then the rate of quenching depends on the polarization angle of
the field. For $(\theta=0)$, i.e the  field and the initial state
are polarized in the same direction , one can notice that the
decay in $|s|$  is smooth and the rate of shrinking is small. As
one increases $\theta=\frac{\pi}{3}$, the coherent vector quenched
quickly and  the rate of  shrinking is larger than the previous
case. Finally, we set the polarized angle of the field
$\theta=\frac{\pi}{2}$, i.e the field and the initial state are in
a normal polarization. In this case $|s|$, decreases faster and
the rate of shrinking is very large.

\begin{figure}
  \begin{center}
\includegraphics[width=25pc,height=15pc]{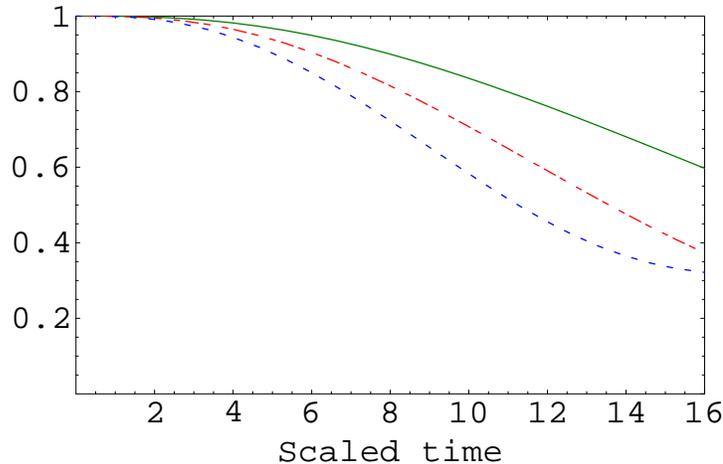}
    \caption{The same as Fig.(3) but
       $\rho{(0)}=\frac{1}{2}(1+\sigma_z)$ with $\theta=\frac{\pi}{4}$.
    The solid, dashed and dot  curves for $\kappa =0.05, 0.07 $ and $0.09$ respectively.}
  \end{center}
\end{figure}

Fig.$(3b)$, shows the dynamics of the amplitude of the coherent
vector for a different class of the initial state, where we assume
that the initial system is prepared in
$\rho(0)=\frac{1}{2}(1+\sigma_y)$. In general the behavior of
$|s|$ is in  agreement with that shown  in Fig.(3a), but for
$\theta=\frac{\pi}{3}$, the decay appears much smoother and
becomes constant earlier than that depicted in Fig.(3a).

The effect of the field parameter is seen in Fig.$(4)$, where we
assume that the initial state of the system is prepared in the
$z-$ direction $\rho(0)=\frac{1}{2}(1+\sigma_z)$ and the field  is
polarized such that  $\theta=\frac{\pi}{4}$.  From this figure it
is clear that the amplitudes value of the coherent vector $|s|=1$
for  a short time $\tau\simeq 2.5$. This mean state the initial
state is robust against the external noise through this time. As
time goes on the initial state turns into a mixed state smoothly
and gradually. Also,  as one increases the values of the field
strength, $|s|$  decreases smoothly and gradually, but the rate of
shrinking increases as one increases $\kappa$.

\section{Fidelity of transmitted Information}

In quantum commuting  context, the  information is transported by
its carrier (states) from one site to another to achieve the final
result. So, it is an important task to preserve  these information
to be used in a further operation \cite{Mur}.
 In this subsection, we quantify the amount of
transferred information between  any two quantum computing
operation. One of the most common measure is the fidelity
$\mathcal{F}$, which measures how much the sended information
related to the initial coded information. Fig.(5) and Fig.(6) show
snapshots of the dynamics of the fidelity for different classes of
initial states  and different polarization of the field.
\begin{figure}
  \begin{center}
    \includegraphics[width=15pc,height=13pc]{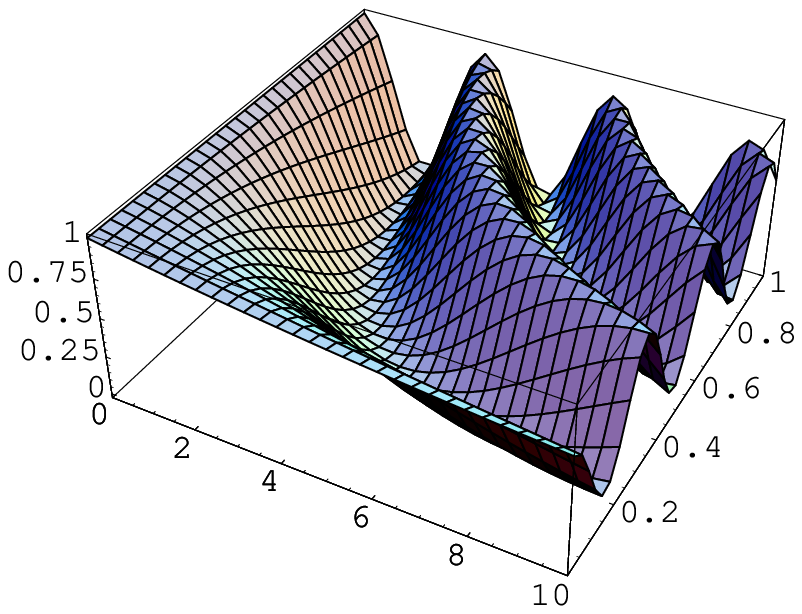}
  \put(-190,80){$\mathcal{F}$}
 \put(-10,48){$\kappa$}
   \put(-160,10){(a)} \put(70,10){(b)}
   \put(-110,20){$\tau$}~\quad
  \includegraphics[width=15pc,height=13pc]{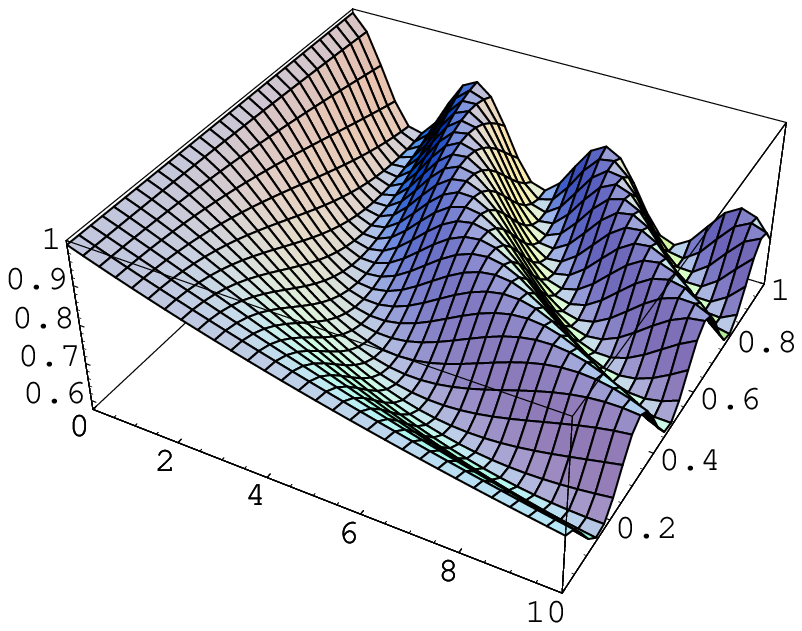}
\put(-190,80){$\mathcal{F}$}
   \put(-10,48){$\kappa $}
   \put(-102,23){$\tau$}
      \caption{The fidelity $\mathcal{F}$ of the  evolved
    density operator, where the initial state  of  the system  is
     $\rho_s(0)=\frac{1}{2}(1+\sigma_x)$. The polarized angle of the field is
     (a)$\theta=0.$ (b) $\theta=\frac{\pi}{3}$. }
  \end{center}
\end{figure}

\begin{figure}
  \begin{center}
 \includegraphics[width=15pc,height=13pc]{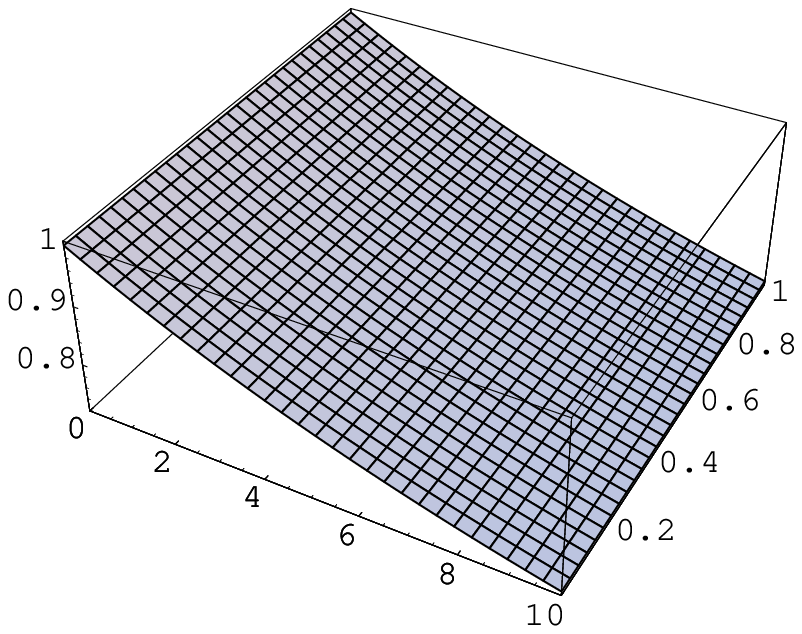}
  \put(-120,15){$\tau$}
  \put(-190,80){$\mathcal{F}$}
 \includegraphics[width=15pc,height=13pc]{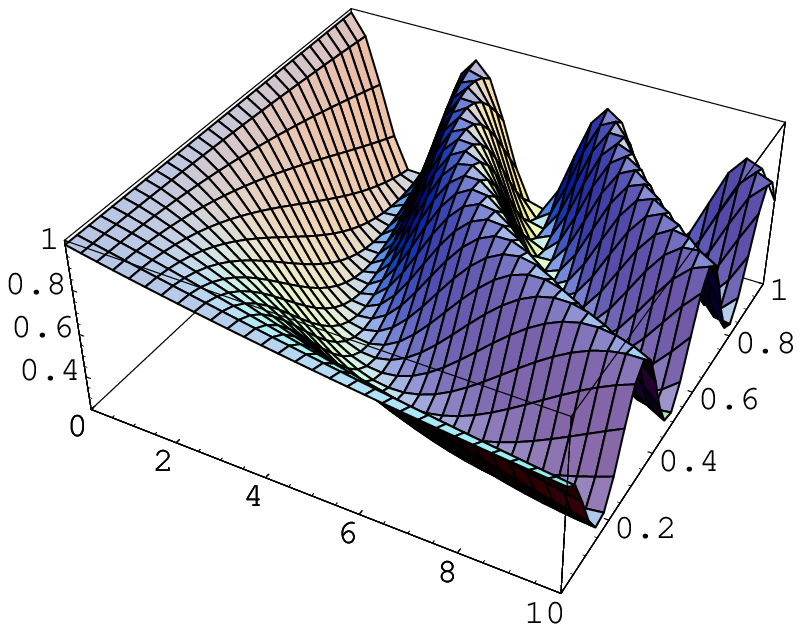}
  \put(-330,0){(a)} \put(-110,-5){(b)}
  \put(-110,15){$\tau$}
     \caption{The same as Fig.(5), but the system is initially
     prepared in $\rho_s(0)=\frac{1}{2}(1+\sigma_y)$. (a)$\theta=0$.
      (b) $\theta=\frac{\pi}{3}$.}
  \end{center}
\end{figure}

\begin{figure}
  \begin{center}
 \includegraphics[width=15pc,height=13pc]{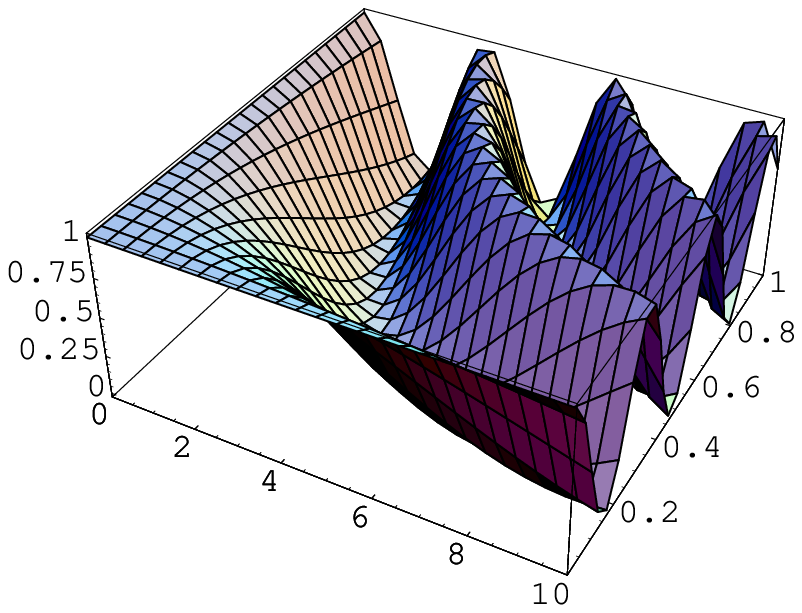}
 \put(-190,80){$\mathcal{F}$}
 \put(-120,15){$\tau$}
 \includegraphics[width=15pc,height=13pc]{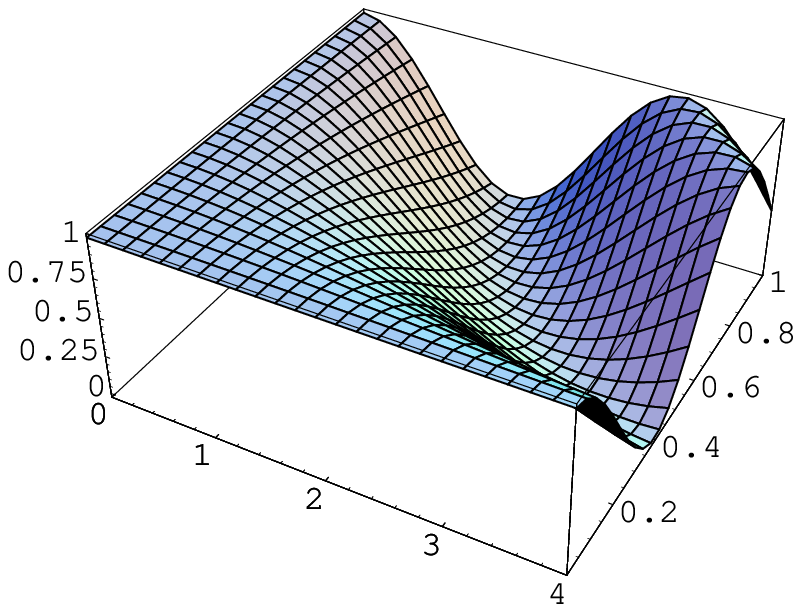}
  \put(-110,15){$\tau$}
  \put(-330,0){(a)} \put(-110,-5){(b)}
     \caption{(a)The same as Fig.(5), but the system is initially
     prepared in $\rho_s(0)=\frac{1}{2}(1+\sigma_z)$.  with $\theta=0$.
      (b) The  same as (a) but for small range of time. }
  \end{center}
\end{figure}

The behavior of the fidelity, $\mathcal{F}$  of the transformed
information for a system is prepared initially  in
$\rho_s(0)=\frac{1}{2}(1+\sigma_x)$ and the field is polarized in
$x-$ axis is shown in Fig.(5a). It is clear that the fidelity
$\mathcal{F}$, oscillates between $1$ at for pure state and $0$
for completely mixed state. Also, as the time goes on the fidelity
decays  smoothly and gradually, but does not reach to zero. This
means that for this initial condition, one can transform the
information between any two operation with non-zero fidelity. Also
as $\kappa$, increases the maximum value of the $\mathcal{F}$ is
smaller than those for small $\kappa$. Fig.(5b), displays the
behavior of the fidelity for different polarization angle
$(\theta=\frac{\pi}{3})$.  In this case the fidelity decreases
faster as the time goes on and  it  completely vanishes  for some
values of $\kappa$ and at specific time. On the other hand the
minimum  values of the rate of information loss increases
 for large values of $\kappa$.

Figs.(6) shows the dynamical behavior of information for different
initial setting. It is evident that, if the system and the field
are prepared initially in different polarization, the fidelity
decays abruptly and very fast. This is clear from Fig.(6a), where
the angle polarization of the field $\theta=0$ and the system is
prepared in $\rho_s(0)=\frac{1}{2}(1+\sigma_y)$. As one increases
the polarization  angle the situation is dramatically changes. As
an example if we set $\theta=\frac{\pi}{3}$, the fidelity
decreases gradually and oscillates between $1$ (for pure states)
and 0 (for completely mixed states.

In the last  example we considered  the initial state of the
system is polarized in the $z-$ axis i.e
$\rho_s(0)=\frac{1}{2}(1+\sigma_z).$ and the polarization angle of
the field $\theta=0$. The dynamics of the fidelity of this system
is shown in Fig.(7a). It is clear that in general the fidelity has
a similar  behavior as that shown in Figs.$(5\&6)$, but the
maximum value of the fidelity is always better. For small interval
of time the fidelity $\mathcal{F}=1$ this is clear from Fig.(7b),
where we plot the fidelity for small range of time. This means
that through this time the information is transformed completely
and correctly. In this case the polarization of the field has a
similar effect if it is polarized in $y-$ axis.

One can conclude that the  polarization  angle of the field has a
devastating impact on the transformed information, if the initial
state and the field are prepared initially in a different
polarization. Conversely, if the field and the initial state are
in a common polarization, then the rate of information loss
increases smoothly and gradually.

\section{Conclusion}
In this contribution, we introduce a complete description of the
transport of information, where we investigated the dynamics of
the purity, entropy and the amplitude of coherent vector.
 These physical quantities represent  the most important properties of
the evolved state through the Kane Solid state computer. Also we
have  shown the effect of the field parameter and the initial
state setting on the amount of the transformed information between
any two operations through the Kane quantum computer.

From our findings, it is evident that the polarization of the
field as well as the initial state setting play central roles  on
the phenomena of the abrupt and gradual decay of the purity and
the amplitude value of the coherent vector. We have shown that, if
one can generate the initial state such that it has the same
polarization of the field, then one can keep the purity of the
evolved state survival for a long time and consequently the rate
of exchange of  information increases slowly and gradually.
Conversely, when both the initial state   and the external field
are polarized in different directions, then the   destructive
influence  on the purity of the carrier of information appears
clearly where the state turns into a completely mixed state
abruptly and very fast.

 The fidelity of the transmitted information is very sensitive to the polarization
of the field. We showed that as one increases the field parameter,
the fidelity decreases and increases but the maximum values always
decreases. On the other hand, for different polarization, the
fidelity decreases abrupt and very fast, while the entropy
increases abruptly. The dynamics of both phenomena changes
gradually if the initial state and the field are polarized in the
same direction. For some classes of initial states, the
information is transformed from a node to another completely and
correctly.


\begin{thebibliography}{0}
\bibitem{shor} P. W. Shor, SIAM J. Comp. 26, 1484 (1997).
\bibitem{Mich} M.A. Nielsen and I. L. Chuang" Quantum computation
and Information"Cambridge Uuniversity press(2000).
\bibitem{Br}
S. L. Braunstein1, C. M. Caves, R. Jozsa, N. Linden, S. Popescu
and R. Schack, Phys. Rev. Lett.{\bf 83} 1054 (1999).

\bibitem{Deut} D. Deutsch, Proc. Roy. Soc. London Ser. A {\bf 400} 97 (1985).

\bibitem{vand} L. M. K. Vandersypen, M. Steffen, G. Breyta, C. D.
Yannoni, M. H. Sherwood and I. L. Chuang, Nature {\bf 414} 46
(2001).

\bibitem{Jones} J. Jones, progress in NMR Spectroscopy {\bf 38}
325 (2001).

\bibitem{Cirac}J.I. Cirac and P. Zoller, Phys. Rev. Lett. 74, 4091
(1995; C. Monroe et al., Phys. Rev. Lett. 75, 4714 (1995).

\bibitem{sor}
 A. Sorensen and K. Molmer, Phys. Rev. Lett. 82, 1971 (1999); A.
Sorensen and K. Molmer, Phys. Rev. A 62, 022311 (2000).

\bibitem{Mak}
Y. Makhlin, G. Sch¨on, and A. Shnirman, Rev. Mod. Phys.{\bf  73},
357 (2001).
\bibitem{sarma}
 V.W. Scarola, S. Das Sarma,  Phys. Rev. A 71, 032340
(2005)

\bibitem{kane} B. E. Kane  Nature, {\bf 393} 1331 (1998).

\bibitem{Bacon}
D. Bacon, J. Kempe, D. A. Lidar and K.B. Whaley, Phys; Rev. Lett
{\bf 85} 1758 (2000).
\bibitem{Well}
C. J. Wellard and L. C. L. Hollenberg, J. Phys. D {\bf 35} 2499
(2002).

\bibitem{Bui} P. V. Buividovich, V. I. Kuvshinov, Phys. Rev. A 73
022336 (2006).
\bibitem{Murt} M. cetinabs and J. Wilkie, J. Phys. A. {\bf 41}
065302 (2008).
\bibitem{zur} W. Zurek, S. Habib and J. Paz, Phys. Rev. Lett {\bf
70} 1187 (1993).
 \bibitem{Jun}  Jun O.S. Yin and  S.J. van Enk, Phys. Rev. A{\bf 77}, 062333
 (2008).
\bibitem{Yan} Y. Xiang and  S.-Jie Xiong, Phys. Rev. A {\bf 76}, 014306
(2007).
\bibitem{Hol} H. F. Hofmann,
 J. Opt. B: Quantum Semiclass. Opt. {\bf 7} S208-S214(2005).

\bibitem{Gen} G. Kimura, Phys. Lett.{\bf A} 314, 339 (2003).

\bibitem{nasser}
B.-G. Englert and N. Metwally, J. Mod. Opt. 47, 221 (2000); B.-G.
Englert and N. Metwally, App. Phys. B72 55(2001).

\bibitem{Guntr}
G. Mahler and V. A. Weberru\ss," Quantum Network:Dynamics of open
nanostructure" second edition (Springer 1998).

%\bibitem{Nmetwally} N. Metwally and M. Abdel-Aty, Int. J. Mod.
%Phys. C, in press(2009).

 \bibitem{Mur} M. Murphy, L. Jiang, N. Khaneja and  T. Calarco
Comments: 4 pages, 4 figures Journal-ref: Phys. Rev. A{\bf 79},
020301(R) (2009).

\end{thebibliography}
\end{document}